\documentclass[11pt]{article}

\usepackage{graphicx,amsmath,amsfonts,amssymb,dcolumn,amsthm,euscript}

\begin{document}

\def\demi{{1\over 2}}
\def\quart{{1\over 4}}
 
\def\dd{d+s}
 \def\ll{\lambda^\t}
\def\d{\delta}
\def\trois{{3\over 2}}

\def\ba{{\bar\alpha}}

\def\ee{\eea}
\def\be{\bea}
 
\def\eq{\eeq}
\def\bq{\beq}
\def\ie{\hbox{\it i.e.}}        
\def\etc{\hbox{\it etc.}}
 \def\eg{\hbox{\ite.g.}}   
  \def\cf{\hbox{\it cf.}}

 \def\etal{\hbox{\it et al.}}
\def\tr{\mathop{\rm tr}}
 \def\Tr{\mathop{\rm Tr}} 
\def\beq{\begin{equation}}
\def\eeq{\end{equation}} 
\def\bea{\begin{eqnarray}}
 \def\eea{\end{eqnarray}}

\def\bar{\overline}
 \def\z{{\bar {z}}}
 \def\nn{\nonumber} 
\def\pa{\partial}
\def\d{{\cal D}} 
\def\I{{\cal I}} 
\def\t{{\theta}} 
\def\s{\sigma}

 \def\pt{\partial_\t}
\def\ic{\int_0^T}

\def\vq{\vec q}

\def\P{\Psi}
\def\bP{\bar \Psi}

\def\vP{\vec \P}
\def\vbP{\vec {\bP}}

\def\l{\lambda}
\def\vl{\vec \l}

\def\eij{\epsilon^{ij}}

\def\V{{\dot q}_i+{{\delta V}\over{\delta q_i}}}
\def\VP{{\dot \P}_i+{{\delta ^2 V}\over{\delta q_i \delta q_j}}\P_j}

\def\et{\epsilon(t)}
\def\Vt{{\dot q}_i+{{\delta \t}\over{\delta q_i}}}
\def\v{{\dot q}_i+f{{\eij q_j}\over{\vec q^2}}}
\def\VPt{{\dot \P}_i+{{\delta ^2 \t}\over{g\delta q_i \delta q_j}}\P_j}
\def\p{\vec p}
\def\bQ{\bar Q}
\def\de{\delta}
 \def\al{\alpha}
\def\ph{E,n}

 \def\demi{{1\over 2}} 
\def\quart{{1\over 4}} 

\title{\textbf{Chains of topological oscillators with instantons \\
   {and  calculable   topological observables in topological quantum mechanics  }}}
\author{\textbf{L.~Baulieu$^\dagger$  and F.~Toppan$^{\dagger \dagger}$  }\thanks{baulieu@lpthe.jussieu.fr, ftoppan@gmail.com
}\\\\
\textit{{ $^{  \dagger}$  LPTHE $-$   Sorbonne Universit\'es, UPMC} }\\
\textit{{ }  4 Place Jussieu, 75005 Paris, France}\\
\textit{{ $^{\dagger \dagger}$  CBPF,   Rio de Janeiro }}\\
\textit{{   Rua Dr. Xavier Sigaud 150, Urca,
cep 22290-180  (RJ), Brazil}}\\
}

\date{}
\maketitle
\begin{centerline} 
\ This work will appear in a special   Nuclear Physics issue "Memoriam Raymond Stora"
\end{centerline}

\begin{abstract}
We extend to a possibly infinite chain the conformally invariant mechanical system that  was 
introduced earlier  as a toy model for understanding the topological  Yang--Mills  theory.
It gives a  topological quantum model that has interesting and  computable  zero modes
 and  topological~observables. 
 It confirms  the recent conjecture by several  authors 
 that supersymmetric quantum mechanics may  provide useful tools for understanding   robotic  mechanical systems (Vitelli et al.) and condensed matter properties (Kane et al.), where  trajectories of effective  models are allowed or not  by the conservation of topological indices.  
  The absences of   ground state and   mass gaps are
special features of such systems. 
\end{abstract}

\section{Introduction}

Topological Quantum Field Theories TQFT's are possible realisations of  the invariance under general local field  transformations general coordinates invariant
symmetries. Such  an invariance    goes  beyond that of current gauge theories.
 The first non-trivial example of a TQFT was introduced by Witten  \cite{wittendona} showing  that the genuine $N=2$ supersymmetric 
 gauge theories   contains observables that describe the Donaldson invariants. The  reinterpretation \cite{bs} of this theory   appeared soon after,  as a suitably gauge-fixed  quantum field theory  stemming from  a classical topological invariant  that explores the BRST cohomology of general Yang--Mills field transformations  modulo ordinary gauge transformations. 
 Because the pattern of   TQFT's is that of an   ultimate type of gauge-fixing,  and because they can be solved, they greatly interested Raymond Stora, who wrote himself a very interesting article on the subject \cite{ OUVRYSTORA}.

At the heart of TQFT's,  the topological  BRST  nilpotent operator $Q$ plays a fundamental role. It  is such that the TQFT Hamiltonian is basically 
$H=\demi[Q,\bar Q]$. One often defines the physical Hilbert space as
the cohomology of $Q$ (states which are annihilated by $Q$ without
being the $Q$ transformation of other states). This unambiguous definition of observables 
  from   the cohomology of a BRST operator   is perfectly suited for the gauge
theories of elementary particles (where the expression of $Q$ is more restricted in comparison to that of a TQFT  and the relation between $Q$ and $H$ is different).
  The cohomology of a TQFT is often contained in another cohomology, in which case it is called an equivalent cohomology~\cite{bs,OUVRYSTORA,review}.
 There were doubts  for a while on the
validity of this construction of  TQFT's, so~\cite{ella} defined and explored a 
solvable  quantum mechanical supersymmetric example to check very precisely all the details and confirm the construction.
The model was that of particle moving in a punctured plane, where  the closed trajectories carry topology because  of their  non-trivial winding numbers.
 Instantons exist  in this case because one  choses in this case a  potential that yields a supersymmetric action 
 with    twisted scale and vector vector supersymmetries, in fact a superconformal supersymmetry. 
 Strikingly, the  ingredients for constructing the model completely reproduce those of   the much more involved Yang--Mills topological theory and we   completely solved it in ~\cite{ella}.    The goal of this article is  to generalise this model to  a more physical multiparticle case with conformal interactions. To do so, we need first to review  some details  of~\cite{ella}.

Afterward we  will show that~\cite{ella} can be extended into a 
very intriguing  model,  which is quite beautiful  and might furthermore  have  richer  applications in practical  domains.
 It   is an explicit example of what was    foreseen long time ago in~\cite{jackiw}, for building some robots with rotational  constrained degrees of freedom, and more recently       by condensed matter physicists, for instance  \cite{vitelli}. Our model  generalises ~\cite{ella} and 
 gives  a  sort of conformally vibrating lattice where each site is a particle  interacting by superconformal interactions with its nearest neighbours (two in this present case). This model exhibits non-trivial  instanton solutions  and has  some topological observables.

\section{The one-particle conformal supersymmetric topological model}
%
%

The model is    a  quantum mechanical system of a particle  moving  in 
  a 2D-plane where one excludes the origin and submitted to a potential we will shortly display.  One  
  has a non-trivial  topological structure because of  trajectories 
with different possible winding numbers $0\leq N\leq \infty$  around the origin. The classical topological symmetry is the group  of arbitrary local  deformations of each particle  trajectory. They can be 
possibly defined modulo  local dilatations of the distance   of the particle to the origin. 
We will see that the model is a  conformal one.

We  call the time by the real variable $t$   and the       
Euclidian time by $\tau$, with $t=i\tau$. The
cartesian coordinates on the plane are $q_i$, with $i=1,2$, and we often use complex coordinates 
$z=q_1+q_2$.

We select
trajectories with periodic conditions.  Namely, the particle  does a closed (multi)-loop 
  between the
initial and final times $t=0$ and $t=T$ (we will choose   $T=1$). 
An integer  winding number  $0\leq N\leq \infty$  is 
assigned to all trajectories  which can be classified   in equivalence classes according  to $N$. 

 As in the Yang-Mills  TQFT
\cite{bs}, one   starts   from a  topological classical action
$\I_{cl}[\vec{q}]$,  (with the above enlarged gauge  symmetry).
It must be  gauge-fixed  in a BRST invariant way  to define a path integral so as one can compute some topological observables.

$\I_{cl}[\vec{q}]$  must be  independent on the time
metric and  on local reformations of trajectories. 
Such  conditions are  satisfied  for
\be \I_{cl}[\vec{q}] = \frac{1}{g}  \int\
 d\t &=& \frac {1}{g} \ic \ d\tau \  \dot\t(\tau)
 = \frac {1}{g}       \ic \ d\tau\ {{\eij \dot
q_i q_j}\over{\vec q^2}}
=\frac {2\pi N}{g}
\ee where
$g$ is a real number that will become a coupling constant. 
In fact, $\I_{cl}[\vec{q}]$
measures the winding number $N$ of the particle (times $\frac {2\pi }{g}$).
It is a  tremendously simplified version  of   the second Chern class $\int d^4x \tr F\wedge F $,   
$F$ being  the curvature of a Yang-Mills field. Here and in what follows
the symbol $\dot X $ denotes ${{d X }\over{d\tau}}$.

The  TQFT path integral is  defined from   a BRST invariant gauge--fixed action 
added to   $\I_{cl}$ as 
 in~\cite{bs}~:
\be \label{pa1}
\int \d [\vec q]\exp -\I_{cl}[\vec{q}]  \to \int \d [\vec q]\exp -\Big (\I_{cl}[\vec{q}]  +{  gauge-fixing} \Big)
\ee
This  way of proceeding  is called nowadays   a localisation procedure.

Once the details of  ``${  gauge-fixing} $"  have  been determined, one can 
  compute topological quantities from  Green functions  of  well chosen  composite operators   ${\it{O}}$.  
\be \label{pa6}
{\rm Topological} \   {\rm information}=\int \d [\vec q]{\it{O}}\exp
\Big (\I_{cl}[\vec{q}]  +{  gauge -fixing} \Big) \neq 0 \ee
The way it goes is as follows.
The  action $\I_{cl}[\vec{q}]$ must be  invariant under the gauge
symmetry
\be\label{topsym}
\vec{q(t)}\to \vec{q(t)}+\vec{\epsilon(t)}\ee 
where $\et$ is any given local shift of the particle position $q(t)$  with appropriate boundary conditions - it cannot  change the winding number of the trajectory. Such shifts can be decomposed in   radial and angular deformations, and one foresees an interesting decomposition between  angular and radial shifts, already noticed in~\cite{ella}.


The BRST transformation laws associated to the symmetry (\ref{topsym})
are basically found by changing the parameters $\epsilon(t) $ of arbitrary shifts into an anticommuting   ghost  $\Psi(t)$ and an anticommuting antighost 
$\bar \Psi(t)$, and introducing 
a Lagrange multiplier $\vl(t)$, with
\be
s\vq=\vP\quad s\vP=0\quad s\vbP=\vl\quad s\vl=0
\ee 
The operation $s$ acts on field functions as a differential
operator graded by the ghost number.
A formal superfield unification exists that  unifies   $\vP(t)$, $\vbP(t)$ and  $ d\vec{q(t)}$ in a single quantity and~$s$~can be interpreted as a differential operator. 
  
To get a gauge-fixed action with a quadratic dependence on the
velocity $\vec{\dot q}$, one chooses  a localisation gauge function   $\V$.
The  prepotential $V[q]$ is a priori  arbitrary, but the  equivariance with respect to dilatations
determines  its dependence on  $\vq$, as it will be shown~shortly.
The gauge-fixing term is $s$-exact, and gives  the  (supersymmetric)  BRST invariant action $\I_{gf}$
\be
\label{IGF} \I_{gf}
[\vq,\vP,\vbP,\vl]
&=&
\ic \ d\tau \left(
\frac {1}{g} \dot\t -s\bP_(
\demi\l_i 
+\V)
\right)
\nn\\
&=&
\ic \ d\tau \left(    \frac {1}{g} \dot\t -\demi\l_i^2
+\l_i\left(\V\right)
-\bP_i\left(\VP\right)
\right)\nn\\
\ee
 Notice that the 
boundary term $\dot V$ adds up to $\dot \t$ after  the elimination of the auxiliary field by its algebraic equation of motion, so that their combination can cancel  out.

The partition function is 
\be
 Z=\int \d [\vq]\d [\vP]\d [\vbP]\d [\vl]\exp -\I_{gf} 
\ee
and, given some functionals ${\it{O}}(\vec q)$  one has well-defined Euclidian path integrals
\be
 <{\it{O}}>=\int \d [\vq]\d [\vP]\d [\vbP]\d [\vl] {\it{O}}\exp -\I_{gf} 
 \ee
A Faddeev--Popov  field theory interpretation of the gauge-fixing can be done by considering     $\V$   as a 
gauge function for the quantum variable $\vq$ and  by inspecting the 
Faddeev--Popov determinant  formally  obtained by    a path integration over the ghosts.  The topological non-triviality of the theory occurs when this determinant has zero modes.

  The BRST
invariance of the field polynomial ${\it{O}}$, if any,  allows one to prove  the topological properties of $ <{\it{O}}>$, that is 
the fact it  only depends on the winding number   $N$ of the trajectories.
Any given topological observable must be first computed in a given topological sector $N$, and one can possibly sum over  $N$, sometimes with a relevant regularisation.  
Our  knowledge of supersymmetric quantum mechanics tells us
that this mean value may depend on the class of the function $V$. What
happens is that in the case of topological field theories, the
Euclidian path integral explores the moduli space of the equation
$\V=0$, as a result of the gauge fixing, which is  non-trivial only for relevant  choices of $V$. The topological observables are defined from the cohomologies of the BRST operator with all possible ghost numbers.

A possible way to select the
prepotential $ V(\vq)$ leading to interesting topological information
has been  investigated in \cite{ab}. One  asks for a larger   invariance
of the action  that  is more restrictive than the
topological BRST symmetry, namely a local version of it, for which the
parameter becomes an affine function of the time, with arbitrary
infinitesimal coefficients. It is a called a vector BRST symmetry. In fact, in our case,  it can be identified with 
the requirement of conformal symmetry.

The ``local" BRST transformations $\delta _l$ are
\be
\delta _l\I_{gf}[\vq,\vP,\vbP,\vl]=0
\ee  
where 
\def\at{\eta(t)}  
\be
\delta l\vq=\at\vP
\quad
\delta _l\vP=0
\quad 
\delta _l\vbP=\at\vl-\dot\at\vq
\quad  
\delta _l\vl=\dot\at\vP 
\ee
and $\at=a+bt$ where $a$ and $b$ are constant anticommuting
parameters. The idea of local BRST symmetry was earlier considered in a paper with Raymond Stora
\cite{STORA}, for the sake of interpreting higher order cocycles which
occurs when solving the anomaly consistency conditions. This symmetry  has been
shown to play a role in topological field theories in \cite{rak}.

 $V$ satisfies the following  constraint due to  this local symmetry,
\cite{ab} \be
{\delta V\over \delta q_i}+q_j\ {\delta\sp 2 V\over \delta q_i\ \de
q_j}=0
\ee
This constraint is solved for $V(\vq)\sim \t$  or  $V(\vq)\sim  \log |\vec q |$     where $\t$ is the angle
such that $z=q_1+iq_2=|\vq|\exp i\t$. We introduce again a multiplicative scale with a real number $g$.
By putting this value of $\V$ in
(\ref{IGF}) and eliminating the Lagrange multiplier $\l$ by its
equation of motion we obtain  (all boundary terms compensate against each other thanks to the choice of $V$):\be \I_{gf} [\vq,\vP,\vbP]=
\ic \ d\tau \left( \demi \dot q_i^2+{ \frac{1}{2g^2\vq^2}}
-\bP_i\left(\VPt\right)
\right)
\ee
One has   
\be
\label{U_t}
{{\delta ^2 \t}\over{\delta q_i \delta q_j}} &=&
{1\over{\vq^2}}
\begin{pmatrix}
-\sin 2\t&
\cos 2\t 
\cr
\cos 2\t&
\sin 2\t  
\end{pmatrix}_{ij}
\nn\\
&=&
{1\over{\vq^2}}\begin{pmatrix}
 \cos  \t&
-\sin\t 
\cr
\sin\t&
\cos\t  
\end{pmatrix}
\begin{pmatrix}
0 &
-1 
\cr
-1&
0  
\end{pmatrix}
\begin{pmatrix}
 \cos  \t&
-\sin\t 
\cr
\sin\t&
\cos\t  
\end{pmatrix}_{ij}
\ee

The signal that the theory truly carries some
topological information is the existence of an   instanton
structure, leading to fermionic zero modes. 

Our gauge-fixing gives  an action  whose bosonic part
is the square of the gauge function or localisation function.  Its 
  Euclidian equations of motion are thus obtained when this gauge function vanishes 
\def\v{{\dot q}_i+ \frac{1}{g}{{\eij q_j}\over{\vec q^2}}}
\def\VPt{{\dot \P}_i +   \frac{1}{g}  {{\delta ^2 \t}\over{\delta q_i\delta q_j}}\P_j}
\be \v=0
\ee
\be
\VPt=0
\ee
In the complex number representation  with  
$z=q_1+iq_2$ and $\P_z=\P_1+i\P_2$, one has  $sz=\P_z$ and one  can write
these equations as
\be
-ig \dot z=    {1 \over{\bar z}} 
\ee
\be 
-ig   \dot \P_z^*=   {1\over{z\z  }}\P_z
\ee
(we use the symbol $^*$ for the complex  conjugation on the ghost  $\P_z^*$ not to do a confusion with antighost $\bar \Psi_z$). Assuming  periodic boundary conditions, one can easily solve the first-order equation for $z$ as a
particle making  $N$ cycles at constant angular speed  with radius $R\sim 1/\sqrt{N}$. More precisely :      \be
2\pi N g- \frac {1}{\z z} =0  \      \ \ \ \ z= \frac{1}{\sqrt{2\pi Ng}} \exp i \theta  \      \ \ \ \  \theta   = 2\pi N t
\ee
Therefore, since the action vanishes for such trajectoires, one finds that the model has  instantons  indexed by an integer $N$,
which are circles described at fixed angular speed with 
  frequency $2\pi N g $, a  tremendously simplified version of the Yang--Mills instantons.


The  equation of motion of the fermion  can be written as
\be
ig\dot\Psi_z^* +\frac {1}{{\z z}  }   U_\theta \Psi _z =0
\ee
\be
 U_\theta   =    R_{\theta}    C  R_{\theta} 
 \ee
 $ U_\theta $ is in fact the subgroup of the large rotation group $O(2)$ elements  connected to the inversion/conjugation matrix $C$ with $\det C=-1$. The other components  of $O(2)$,  the   $SO(2)$ ``small" rotation matrices $R$, with $\det R=1 $,  are  connected to the identity. 
  In matrix notations,  one has for the vector representation :
 \be
 C=
 \begin{pmatrix}
0 &-1 
\cr
-1&
0  
\end{pmatrix}
\ee
Thus the meaning of Eq.~(\ref {U_t}) is 
 \be
   \frac {1}{{\z z}  }      \Big (U_\theta  \Big )  _{ij}    &=&{{\delta ^2 \t}\over{\delta q_i \delta q_j}} 
  = {{\delta ^2 Log  \sqrt {z\z }}\over{\delta q_i \delta q_j}} 
=\frac{1}{{\vec q}^2}
R_\t C
R_\t
\ee
In the  complex number   representation   
  the rotation $R_{ \t}$ is   $z\to \exp i\t  \ z$ and the matrix $U_\t$ is  the complex conjugation followed by  the multiplication by   $\exp 2i  \t  $, $z\to \exp 2i\t \  \bar z$.

The solution of  the equation of  fermionic  zero mode in the field of the instanton with winding number $N$  is obtained by defining 
\be
\Psi (t)  =    \Psi_0\exp - i2\pi Nt 
\ee
that implies that    $\Psi_0$  is time independent and  satisfies 
\be
  \Psi^*_0 -  C  \Psi _0 =0
\ee
Thus  $\Psi _0 $ is   an eigen-vector of the operator $C$.
 Depending on the orientation of the winding number $N$ of the instanton, we have a  fermionic zero mode  
\be
\Psi _0
=
 \begin{pmatrix}
1 
\cr
\pm1 
    \end{pmatrix}
\ee
 
The fermionic zero modes can be drawn as     constant vectors     attached to  the particle and  running  along     the circles of radius $\sqrt {2\pi Ng}$  at  a constant  angular speed  $2\pi Ng$.

To summarise,  for each instanton (here we  reinstall the period $T$ previously rescaled at $T=1$), 
\be
z^{(n)}=\sqrt{T\over{2\pi N }}\exp -i{{2\pi Nt}\over T} \quad\quad N \in Z
\ee
one has a ghost zero mode
\be
\P_z^{(n)}=\Psi_{0z}\exp  - i {{2\pi Nt}\over T}
\ee
where $\Psi_{0z}$ is a constant   Grassmann variable. The Euclidian energy and angular
momentum of the action evaluated for these field configurations vanish
for all values of~$N$.
 
Because of these degenerate zero modes,    
 BRST-exact observables  exist  with non-vanishing 
mean values. They are in fact   independent  on  energy and time
because of the BRST-exactness of the action. 
  Moreover, they can be computed as   a series over the topological  sector index $N$.

We verified  these properties in ~\cite{ella}  by explicit computation in Hamiltonian formalism,   using the technics developed in~\cite{fubini},\cite{rafu},\cite{macfarlane},\cite{Comtet}.
 
 We found a continuum spectrum of   states that are normalisable as plane waves in one dimension. This is
in fact a consequence of the continuity of the spectrum  of the Hamiltonian in the radial
direction. They build an appropriate basis of stationary solutions
since, with the appropriate  normalisations factor,
one has $\sum_n\int _{E>0}dE |E,n><E,n|=1$, where $n$ is the angular momentum quantum number.  On the other hand, for
$E=0$, the solvable  Schr\"odinger has  
  no admissible normalisable solution.  Thus we have a continuum
spectrum, bounded from below, with a ``spin" degeneracy equal to $4$ and
an infinite degeneracy in  $n$. The
peculiarity of this spectrum is that there is no ground state, since
we have states with energy as little as we want, but we cannot have
$E=0$. This is a consequence of the conformal property of the
potential $1\over{|\vq|^2}$.

Since we cannot reach the energy zero which would be the only $Q$ and
$\bar Q$ invariant state, we concluded that supersymmetry is broken, and that the model is for a non-trivial 
topological supersymmetric quantum mechanics.

As for the computation of BRST invariant observables, we found by  dimensional arguments 
that the candidates are  (in polar coordinates) \be
{\it{O}}_\t=[Q,r\bP_\t]_+ =[\bQ,r\P_\t]^\dagger_+
\quad\quad
{\it{O}}_r=[Q,r\bP_r]_+=[\bQ,r\P_r]^\dagger_+
\ee
The   mean values of these operators
between normalised states  could be computed as 
\def\B{J_{\sqrt{n^2+f^2}}}
\be\label{obs}
{{<E,n| [Q,r\bP_\t]_+|E,n>}\over{<E,n|E,n>}}=n+i\frac{1}{g}
\ee
and
\be\label{nobs}
{{<E,n| [Q,r\bP_r]_+|E,n>}\over{<E,n|E,n>}}=\lim_{L\to\infty}
{{L^2\B^2(L)}\over{\int_0^L dr\B(r)}} 
\ee
The last quantity is  bounded but ill-defined, so we rejected it.  Therefore,  for any normalised state $|\phi_n>=\int dE
\rho(E)|E,n>$ with a given angular momentum $n$, the expectation value
of $[Q,r\bP_\t]_+ $ is
\be\label{obs4}
<\phi_n|[Q,r\bP_\t]_+|\phi_n>=n+i\frac{1}{g}
\ee  
independently of the weighting function $\rho$.

If we now sum over all values of $n$, what remains is the topological
number
\be\label{obs1}
<{\it{O}}_\t>=\sum_n <\phi_n|[Q,r\bP_\t]_+|\phi_n>=
\sum_n n +i\frac{1}{g}\sum_n1
\ee

Our  result \cite{ella} meant that there are two
observables, organized in a complex form, in the cohomology of the
punctured plane.

	\section     {    What' new after years : Chains of topological oscillators  with conformal properties            }

We wish  a system  of equations for the interactions  of particles
confined in  a 2D-surface with  planar complex coordinates    $  z_{n}    $   and   possible conformal interactions between next neighbours as a generalisation of what we presented  in the previous section. Non-trivial topology can arise because the potential is such that particles cannot sit on top of each other.

 For concrete purposes and the sake of simplicity,   we ask that the particles  can have stable or metastable     rest solutions on a line  at given locations  $u_n $, with   $\frac{ du_n}{dt}=0$.  
We    choose the following particle alternate rest positions on a line
\be    
 u_{2n}   &=&     2n a    \nn \\
 u  _         {2n+1}    &=&     2na+b      
   \ee   	
   In fact one has
	\be
	u_{2n+1  } -u_{2n} &=&  b    \nn \\
	u_{2n+2  } -u_{2n+1} &=&      2a-b    \equiv c   \nn \\
	u_{2n+3  } -u_{2n+2  }  &=& 2na+2a +b-  2na -2a=b\nn\\
	\etc\ldots
	\ee
   So the distance   between the site  $ u_{2n}$  and its  left  neighbour  $ u  _         {2n-1}$  is    $  b   $ and between $u_n$ and  its right neighbour  $ u  _         {2n+1}$    it is  $c= 2a-b$, $b\neq c$,  and so on 
	\be
	\ldots  
	 \rightarrow  (  u_{2n-1}) \leftarrow b  \rightarrow 
	 (  u_{2n})  \leftarrow c  \rightarrow   
		(  u_{2n+1  })  \leftarrow  b \rightarrow 
	(  u_{2n+2})  \leftarrow    c  \rightarrow   
	(  u_{2n+3})  \leftarrow  b  \rightarrow 
	\ldots
	\nn \\
	\ee
	In fact, we wish  to build a system that is analogous to a  bidimensional crystal of particles that interacts with few (here two) of 	their neighbours, by demanding  some conformal properties and  instanton solutions. The use of supersymmetric quantum mechanics is thus desirable to define and compute  topological  invariants for they system   by  path integration.
	
	We have in mind  to describe,  in particular, systems as in   \cite{jackiw}, and recently,  \cite{vitelli}, for rotor models,  chemical chains, etc...
	
	The potential  we will introduce will not give an integrable model. Rather, it has  classical solutions that   reproduce the behaviour  systems with  rigid links between points,  such as   articulate bars with   rotation freedom, for instance  the   rods   coupling the wheels of a  locomotive,  and the parameters can be adjusted such one has  a global movement,  with a careful adjustments  of the articulations for which some indices don't vanish.
	
	In fact, the multivalued prepotential  we will introduce is  just what is needed to possibly go  ``off-shell" 
	from the classical behaviour of an articulated   classical system,  with stable  classical trajectories corresponding  to rigid links  between its elements.
	 One builds a supersymmetric model  to calculate    indices and/or  topological numbers that ensures non-trivial propagations, such as wave packets that are soliton  and   spin-like waves. The power of a TQFT is that, when  one does the path integration and  when the bosonic part of the  classical systems hits a   given instanton,    fermionic zero modes can occur, and their path integration contributes by a normalised amount to a topological observables. 	In fact the Yang--Mills TQFT works this way, but our model just does the same in a much more concrete way.

 We   will be also concerned on possible limits, when $n$ can be replaced by a continuous variable~$x$, as standard way of dealing with    a very large number of particles, with a $1+1$ field theory limit.
		
		\subsection {The equations}
		\def\Z{   {\bar Z}}
		It is not a restriction to consider a generic value of $n$ that is even. By inspiration from the Yang--Mills self-dual equations and the toy model \cite{ella}, we choose
		\be
		ig \dot \z_n    &=&    \frac  {1}{     z_n- z_{n-1}-b  }   +   \frac  {1}{     z_{n }- z_{n+1} +c }  \nn\\
		ig \dot \z_{n+1 }    &=&    \frac  {1}{     z_{n+1}  - z_{n}-c }   +   \frac  {1}{     z_{n+1}- z_{n+2} +b } 
		\ee
		So we have
		\be
		 ig(   \dot \z_{n+1 }  - \dot \z_n  ) &=&    \frac  {2}{     z_{n+1}  - z_{n}-c }   -  \frac  {1}{     z_{n+2}- z_{n+1} -b } - \frac  {1}{     z_n- z_{n-1}-b  }
		\ee
		We find indeed that, for these equations,  we have  static solutions for $z_p=u_p$, with \  $\dot u_p=0$, for all values of $p$.
		
		If we define the $Z$'s as 
		\be
		z_n(t) =u_n +Z_n (t)   \ee
		we have
		 \be\label{eqloc}
		ig \dot \Z_n    &=&    \frac  {1}{     Z_n- Z_{n-1}  }   +   \frac  {1}{     Z_{n }- Z_{n+1}   }  \nn\\
		\ee
		with 
		\be
		 ig(   \dot \Z_{n+1 }  - \dot \Z_n  ) &=&    \frac  {2}{     Z_{n+1}  - Z_{n} }   -  \frac  {1}{     Z_{n+2}- Z_{n+1}   } - \frac  {1}{     Z_n- Z _{n-1}  }
		\ee
		These equations  give solutions to the equations of motion derived from the topological gauge-fixing of  a  topological term $\int_\Gamma  dV _{{2-\rm {neighbours}}}$,   where  the  multivalued   prepotential $V _{{2-\rm {neighbours}}}$~is 
		 \be
	    V _{{2-\rm {neighbours}}} (   \{    z_n  \}) =\sum_{p  \subset { {\cal  \large Z} }}   \Big (
Log |  z_{2p+1}  - z_{2p}-c | +  Log |  z_{2p }  - z_{2p-1}-b  |   \Big )
\ee
These solutions  extremise the following bosonic action
	\be
	 \int  _\Gamma dt     \sum _{i,p} 
	  \Big( 
	   \frac{1}{2} ( {{{\dot q}^i}_p }  )^2   + 
	\frac{1}{2g^2} \frac{1 }{     (q_p- q_{p-1})^2 }  + \frac{1}{2g^2} \frac{1 }{     (q_p- q_{p+1})^2}
	 \Big )
	 +\int_\Gamma  dV _{{2-\rm {neighbours}}}
	 \ee
		
Some of the solutions of these coupled non-linear equations   are in particular epicycles that are shifted along  successive points on a lines. They are instantons, since they give the same value $S=0$ for   the action.
 In the next  section we indeed  give particular solutions. 	
Generalising the steps for the one particle case the supersymmetric action will be
 \be
	 \int  _\Gamma dt     \sum _{i,p}
	  \Big( 
	   \frac{1}{2} ( {{{\dot q}^i}_p }  )^2   +  \frac{1}{2}  \Big (\frac{\delta V _{{2-\rm {neighbours}}}}    { \delta {{\dot q}^i}_p} \Big)^2
	 \Big )
	 -\sum _{i,j,p,q}
	 \int_\Gamma      \bar \Psi^i_p  \Big ( \delta_{ij}   \delta^{pq}   \frac {d}{d   t} +
	 \frac
	  {\delta ^2
	   V _{{2-\rm {neighbours}}}
	   }   {\delta q^i_p \delta q^j_q}\Big ) \Psi^j_q 
	   \nn\\
	 \ee

We might  extend the computation for more than two next-neighbour oscillators using more length scales
\be
	    V _{4\ neighbours} (   \{    z_n  \}) =\sum_{p  \subset { {\cal  \large Z} }}   \Big (
Log |  z_{2p+1}  - z_{2p}-c | +  Log |  z_{2p }  - z_{2p-1}-b  | \nn\\
+  Log |  z_{2p+2}  - z_{2p}-e | +  Log |  z_{2p }  - z_{2p-2}-f  |    \Big )
\ee
%
%

	\subsection{Epicycles as particular solutions  }
	Let us define
	\be
\Delta_n =   Z_{n}  -  Z_{n-1}  
\ee
Eq.~(\ref{eqloc}) reads as 
\be
		 ig \dot{ \bar  \Delta  }   _ {2p}   & =&
		   \frac  {2}{    \Delta  _{2p} }   -  \frac  {1}{     \Delta   _{2p+1}   } - \frac  {1}{     \Delta   _{2p-1}    }
		   \nn\cr
		    ig \dot{ \bar  \Delta }    _ {2p+1}   & =&
		   \frac  {2}{    \Delta   _{2p+1} }   -  \frac  {1}{     \Delta   _{2p+2}   } - \frac  {1}{     \Delta   _{2p}    }
		\ee
		and we have no need to distinguish between even and odd sites.
		\subsubsection{Solutions with constant frequencies and radii }
		If we take solutions with the same radii $|Z_p|=R$, all particles rotate at the constant angular  speed $\omega
		=\sqrt {2\pi Ng}$.  The  $Z_p$ are the  summits of a regular rotating  regular  polyedra in a circle, and one gets   a representation of   a fixed solid that rotates with a speed related to its dimension. As seen in the rest frame of one of the $z_p$, the particles describe   epicycles. 
		The solution is given by :
		\be
		\Delta _n
		=\frac{1}{\sqrt {2\pi N g} }\exp i(2\pi N  g  t +\delta_n)
		\ee
		and the phases $\delta_n$ satisfy
		\be
		\label{phase}
		\exp i \delta_n = 2 \exp i \delta_{n}-  \exp i \delta_{n-1}- \exp i \delta_{n+1}
		\ee
		that is 
		\be
		\label{phase}
		  \exp i \delta_{n}-  \exp i \delta_{n-1}- \exp i \delta_{n+1}=0 
		\ee
		This equation can be solved using  determinant techniques.

		\subsubsection{Solutions with alternate frequencies and radii}
		We have solutions where the last two off-diagonal terms  on the right hand-side  compensate each other. They are  such that $\Delta_n$ describes circles at a constant  frequency $N^{even}$ for   even  $ n$   and at  a possibly different frequency  $N^{odd}$ for  odd  $n$:
		the phase of   $Z_n(t)$ differs from that of $Z_{n+2}(t)$ by odd numbers of $\pi$, namely
		\be
		\Delta_{2p}&=&    \frac{1} {\sqrt {2\pi N^{even}g}} \exp i (     2\pi N^{even} t   +\delta _{even})  \nn\\
		\Delta_{2p+2}&=&-\frac{1} {\sqrt {2\pi N^{even}g}} \exp i  (  2\pi N^{even} t    +\delta _{even}) \nn\\
		\ee
	and 
	\be
		\Delta_{2p+1}&=& \frac{1} { \sqrt {2\pi N^{odd}g} }
		 \exp i ( 2\pi N^{odd} t        +\delta _{odd})  \nn\\
		\Delta_{2p+3}&=&- \frac{1} { \sqrt {2\pi N^{odd}g} }\exp( 2\pi N^{odd} t        +\delta _{odd})
		\ee  The  radii  are respectively   $  \frac{1} {\sqrt {2\pi N^{even}g}} $ and   $\frac{1} {\sqrt {2\pi N^{odd}g}} $ for the even and  odds $\Delta_n$.
%
%
		
		Going back to the variables $Z$' and $z$ one sees moreover that the  coordinates $z_n$ describe  involved epicycles.

		\subsubsection{The limiting case for     the  wheel rods of a steam locomotive  } 
		There is   interesting limit of the last case, when the frequency of for instance  the particles with odd indices is very large,  $N^{odd}\to \infty$. Then $Z_{2p-1}$  runs  at  a very high speed on a  circles of radius almost equal to zero around  
		$Z_{2p}$, and Eq~(\ref{eqloc})  becomes
		\be\label{eqloc1}
		ig \dot \Z_{2p}    =
		  \frac  {2}{     Z_{2p}  }    \nn\\
		\ee
		modulo terms of order $O(1/N^{odd}g)$. In the limit, one has 
		\be \label{eqloc2}
		z_{2p} & \sim &  2pa +  \frac {1}{\sqrt {4\pi N^{even}g}}  \exp i  ({4\pi N^{even}t +\delta _{even}}) \nn\cr
		z_{2p+2} & \sim &  (2pa+2a)  -\frac {1}{\sqrt {4\pi N^{even}g}}  \exp i({4\pi N^{even}t +\delta _{even}} )
		\ee
Each particle with an odd index  runs on a very  short radius trajectory at a very high frequency, that  is, it is   basically glued by an almost rigid bar  to a particle  with an even index that runs itself on  a circle at finite frequency with  a finite radius     $ \frac {1}{\sqrt {4\pi N^{even}}} $,  centred on a position $u_2p$, which is  almost exactly   the centre of rotation  of a very fast running particle    in a $1\over  _{N^{odd}}$ approximation, and so on.  In   the limit where the winding number $N^{odd}$ goes to infinite, the odd particle  becomes invisible, as a small amplitude local vibration in the middle of a bar ! On the other hand the trajectories  of the even particles  are circles
of radius  $ \sqrt {2\pi   N^{even}g}$        around the points $u_{2n}=2na$, with alternative phase as read in Eq.~(\ref{eqloc2}). For these trajectories one can put a rigid articulate bar between $z_{2n}$ and $z_{2n+2}$.

This interesting solution  describes  the behaviour of the rods of the    wheels of a steam locomotive. The $z_p$ are nothing else than the points where the  rods connect two neighbours wheels. This also describes the apparatus of~\cite{vitelli}.

This is an interesting model where a conformal  potential reproduces on-shell  articulate mechanical attachments.
Since we have a non-linear equations, there are certainly other solutions, with more complicated propagating waves.

 \subsection {The supersymmetric Lagrangian}
%
%
%
%
It is
\be\label{Lsusy}
S
= \int  _\Gamma dt     \sum _p  \frac{1}{2} ( {{{\dot q}^i}_p  } )^2   +  \frac{1} {2} \Big( \frac  {\delta V }   { \delta {q_p^i}   }\Big)^2
-   \sum _{p, r} \    \bar \psi _{p}^i\Big  ( \delta _{p r}  \delta _{ij}   \frac{d}{dt}     + \frac  {\delta^2 V }   { \delta {q_p^i}   \delta {q_r^j}  }          \Big )\psi _{r}^j
\nn\\
\ee
that is, modulo the introduction of auxiliary fields and their elimination from the action
\be
S
= \int _\Gamma dt  \ ,   Q \{ \sum _p     \  \bar 
\psi_p ^i
\Big(
\frac{1}{2} b^i+ 
 {\dot q}_p ^i  +    \frac  {\delta V }   { \delta {q_p^i}   } \Big  ) \} -\int _\Gamma  dV
 \ee
Here the (topological) supersymmetric  graded differential operator is defined as 
\be
Qq = \psi, \ \   
Q \psi =0 ,\ \  
Q\bar\psi  = b,\ \  
Qb = 0
\ee
The   Lagrangian is $Q$-exact, modulo a topological term. The nilpotency of $Q$ proves the $Q$-invariance. 
It is noteworthy that  $S$  is also $\bar Q$-invariant (and $\bar Q$-exact),  modulo a boundary  term, where the  definition  of  $\bar Q$  is  obtained from  that of $Q$ by exchanging $\psi$ and $\bar \psi$ and $b$ and $-b$,
$\bar Q q = \bar \psi, \    
\bar Q \bar \psi =0 ,\   
\bar Q \psi  = b,\   
\bar Q b = 0$.  $Q$ and $\bar Q$ anticommute. The action can be written as a $\bar Q$-exact term. However, the action is not  $Q\bar Q$-exact.

With our choice where $V$ only depends on linear combinations of  $\log |\vec q_p - \vec q_r|$, the action is (super) conformally invariant 
and, moreover, the boundary term $\int dV$ is non-trivial, and discretely multi-valued.

\subsection {Zero modes for the chain}

Let us come back to the  equation   whose solutions extremise the bosonic part of the action  of  our conformal oscillator chain lattice 
\be
		 ig(   \dot \Z_{n+1 }  - \dot \Z_n  ) &=&    \frac  {2}{     Z_{n+1}  - Z_{n} }   -  \frac  {1}{     Z_{n+2}- Z_{n+1}   } - \frac  {1}{     Z_n- Z _{n-1}  }
		\ee
We defined 
$
\Delta _n = Z_{n+1 }  -   Z_n  
 $ and 
the  topological symmetry operation on the particles is    
$
Qz_n=Q Z_n =\psi _n
$.  So by defining 
$
\Psi _n \equiv \psi_{n+1}-\psi_n
$
we have 
\be
QZ_n=\Psi_n
\ee
The supersymmetric Lagrangian is 
\be
Q     \sum_n    \bar\Psi_n  \Big (
 ig  \dot \Delta ^* _{n }      -    \frac  {2}{     \Delta_{n} }   + \frac  {1}{     \Delta_{n+1}   } + \frac  {1}{     \Delta _{n-1}  }
 \Big )
 \ee
Therefore the zero modes for the fermions are the non-trivial solution the $Q$ variation of the topological gauge function when the positions satisfies it, that is \be
ig \dot \Psi  ^* _{n }      -    \frac  {2 U_{\t _{n} } \Psi_n }{   | \Delta_{n} |^2}   + \frac  { U_{\t _{n+1}  }{\Psi_{n +1}}}{    |  \Delta_{n+1} |^2  } 
 +\frac  { U_{\t _{n-1}  }{\Psi_{n -1}}}{    |  \Delta_{n-1} |^2  }  =0
\ee
 Consider the previously found  trajectories  
\be 
\Delta _n = \frac{1}{\sqrt {    2\pi N g}} exp \   (i2\pi Nt +\delta_n)
\ee
There are zero modes
\be
\Psi_n  (t)  =    \Psi_{n,0}\ exp -  i2\pi Nt 
 \ee
where the $\Psi_{n,0}$ are time independent.
Indeed, after   relevant manipulations using the definition of  the matrices $U_\t$, one gets the following conditions :
\be
   \Psi   _{n ,0}      =    2  (\exp 2i\delta_{n} ) \Psi_{n,0}    - (\exp 2 i\delta_{n+1} ) \Psi_{n+1,0} 
-  (\exp 2 i\delta_{n-1})   \Psi_{n-1,0} 
\ee
that can be solved using determinant techniques.
%

The topological invariant are Green-functions of the type
\be
\int \Pi [dz_p] [d\psi_p][d\bar \psi_p][db_p] {\cal{ P} }(\psi, z)
\ee
 where   ${\cal{ P} }$ is $Q$ invariant and has odd fermionic number.
 
We can compute as in \cite{ella}   these invariants in the case of one particle as well of the Witten index of the theory, using the Hamiltonian formalism.
 
 \section{The continuous limit   $n \to \infty $ and $b,c,a \to 0$
 } 
 
 \subsection{The bosonic sector}
 
 We can consider the limit where the index $n$ becomes a continuous one, and we have a chain of sites that can exhibits propagation of non-trivial waves.
 
 This limit is  quite strong:  it  is  not only the difference between the radii of     $\Delta_n$  and $\Delta_{n+1}$ that  becomes small;    the difference between  the angles  $\t (\Delta_n)$  and $\t (\Delta_{n+1})$ is also small. We could consider more subtle limits.
     
Let us suppose that we have a solution for each site
\be
\Delta_n = R \exp  i \theta_n(t) 
\ee
where   $R$ is time independent but $\theta_n(t) $ has a non-trivial $t$ dependence.
Thus the equation of motion of   $ \theta _n $  is              
\be
 -R g   \dot \theta_n   \exp -i \theta_n(t) = \frac {1}{R}( 2  \exp -i  \theta_n  -   \exp -i \theta_{n+1}   -  \exp -i \theta_{n-1}  )\nn\\  
    = \frac { \exp -i  \theta_n}{R} [2    -   \exp -i ( \theta_{n+1}-\theta_n)     -  \exp -i (\theta_{n-1} -\theta_n) ] 
\ee
In the continuous limit, $n$ is replaced by a suitably normalised  continuous length variable $n\to x$, $\theta_n  \to  \t(x)   $, and one can write
\be
i  gR^2 \dot  \theta(x,t)    \sim \frac    { \partial^2 }{\partial^2 x}   \theta(x,t)  +  (  \frac    { \partial }{\partial x}   \theta(x,t)   )^2
\ee
If we change variables,  $\theta =\log v(x,t)$, the former equation reads 
\be
i  gR^2 \dot  v(x,t)    \sim \frac    { \partial^2 }{\partial^2 x}   v(x,t)   
\ee
that is, a   Schr\"odinger type free equations, with periodic boundary conditions  for $\theta(x,t)$. In what follows we replace $\sim$ by an equality.

The general solution is easy to find by   Fourier transform, using the time periodicity of  $v$, 
\be
v(x,t)   =\sum _{N}   v_{N}    \  \exp -i(   2\pi N t-  R\sqrt{2\pi Ng} \  x)     )
\ee
  \be
\t (x,t)   =\exp\sum _{N}   v_{N}    \  \exp -i(   2\pi N t-  R\sqrt{2\pi Ng} \  x)     )
\ee
  
Finally one gets non-trivial wave packets
 \be
\Delta(x)  =R \exp2i\pi   \exp\sum _{N}   v_{N}    \  \exp -i(   2\pi N t-  R\sqrt{2\pi Ng} \  x)     )
\ee


 \subsection{The fermionic zero modes}
 The zero modes continuous field is $\Psi_n\to \Psi(x,t)$. The continuous limit of  
   \be
ig  \dot \Psi  ^* _{n }      -    \frac  {2 U_{\t _{n} } \Psi_n }{   | \Delta_{n} |^2}   + \frac  { U_{\t _{n+1}  }{\Psi_{n +1}}}{    |  \Delta_{n+1} |^2  } 
 +\frac  { U_{\t _{n-1}  }{\Psi_{n -1}}}{    |  \Delta_{n-1} |^2  }  =0
\ee
is \be
 ig \dot \Psi  ^* _{n }    = 
 \frac    { \partial^2 }{\partial^2 x} \Big( \exp2i \theta(x,t) \Psi^* (x,t)\Big) 
  +  
 [  \frac    { \partial }{\partial x} \Big( \exp2i \theta(x,t) \Psi^* (x,t)\Big) ]^2 
   \ee
   or 
\be
ig  \dot \Psi  ^* _{n }    = 
 \frac    { \partial^2 }{\partial^2 x} \Big(v^2(x,t) \Psi ^* (x,t)\Big) 
  +  
 [  \frac    { \partial }{\partial x} \Big(v^2\theta(x,t) \Psi^* (x,t)\Big) ]^2 
   \ee
In short there are zero modes. Most  on them will published in a more extended publication~\cite{toppan}, as well as   more sophisticated  applications of the model.

 \section{Discussion}
We have shown a multi-particle example for which the requirement of local BRST
symmetry   selects a
superconformal quantum mechanical system with intriguing non-linear equations.  It generalises the  more elementary one particle model~\cite{ella} and seems to  provide  a model with apparently deeper applications. The spectrum
of the theory has no ground state and a supersymmetry breaking
mechanism occurs  without the   presence of a dimensionful
parameter.    The generalisation of these observations to a 2d quantum field
theory by the continuous limit we sketched here is an interesting   question.

\vskip .5cm
\noindent {\it \bf Acknowledgements :
 }  
 
  I thank my collaborator F. Toppan who introduced me to the work \cite{vitelli}. This rebooted my interest in the model  \cite{ella} that I have     generalised for the book in the memory of Raymond Stora.  Much more on the properties of this non-linear model will be presented in a joint publication with  F. Toppan.

\end{document}